\documentclass[showpacs,aps,prd,reprint,superscriptaddress,nofootinbib,longbibliography]{revtex4-1}
\usepackage[colorlinks=true, pdfstartview=FitV, linkcolor=magenta,citecolor=blue, urlcolor=magenta,
bookmarks=true, bookmarksnumbered=true, breaklinks]{hyperref}
\usepackage[dvipdfmx]{graphicx}
\usepackage{amsmath,amssymb,amsthm,bm,mathrsfs,tabularx}
\allowdisplaybreaks[3]
\newcommand{\Slash}[1]{{\ooalign{\hfil/\hfil\crcr$#1$}}}

\begin{document}
\title{QCD Kondo excitons}

\author{Daiki~Suenaga}
\email[]{\tt suenaga@mail.ccnu.edu.cn}
\affiliation{Key Laboratory of Quark and Lepton Physics (MOE) and Institute of Particle Physics, Central China Normal University, Wuhan 430079, China}
\author{Kei~Suzuki}
\email[]{{\tt k.suzuki.2010@th.phys.titech.ac.jp}}
\affiliation{Advanced Science Research Center, Japan Atomic Energy Agency (JAEA), Tokai 319-1195, Japan}
\author{Shigehiro~Yasui}
\email[]{\tt yasuis@keio.jp}
\affiliation{Research and Education Center for Natural Sciences,\\ Keio University, Hiyoshi 4-1-1, Yokohama, Kanagawa 223-8521, Japan}

\begin{abstract}
The quantum chromodynamics (QCD) Kondo effect is a quantum phenomenon in which heavy quarks ($c$, $b$) exist as impurity particles in quark matter composed of light quarks ($u$, $d$, $s$) at extremely high density.
This is analogous to the famous Kondo effect in condensed matter physics.
In the present paper, we show theoretically the existence of the ``QCD Kondo excitons", i.e., the bound states of light quarks and  heavy quarks, as the lowest-excitation modes above the ground state of the quark matter governed by the QCD Kondo effect.
These are neutral for color and electric charges, similarly to the Kondo excitons in condensed matter, and they are new type of quasiparticles absent in the normal phase of quark matter. 
The QCD Kondo excitons have various masses and quantum numbers, i.e., flavors and spin parities (scalar, pseudoscalar, vector, and axialvector).
The QCD Kondo excitons lead to the emergence of the neutral currents in transport phenomena, which are measurable in lattice QCD simulations.
The study of the QCD Kondo excitons will provide us with understanding of new universal properties shared by quark matter and condensed matter.
\end{abstract}

\maketitle

\section{Introduction}
\label{sec:Introduction}

The Kondo effect is caused by a strong coupling between an itinerant electron and a spin impurity in metal, which leads to the enhancement of the electric resistance in the low-temperature region, and it still provides us with universal problems in various systems with a non-Abelian interaction like spin exchange~\cite{Kondo:1964} (see also Refs.~\cite{Hewson,Yosida,Yamada,coleman_2015}).
Kondo insulators (Kondo lattices) are composed of bound states (quasiparticles) as superpositions of itinerant electrons and spin impurities, the interaction of which is dynamically enhanced by the Kondo effect~\cite{Kasuya1956,Mott1974,Doniach1977,Lacroix:1979,Coleman1983,Coleman1984,Coleman1985,ReadNewns1984,AuerbachLevin1986,Coleman1987,MillisLee1987,Kasuya1995,OharaHanzawaYosida1997,OharaHanzawaYosida1999,SenthilSachdevVojta2003,OharaHanzawa2009,OharaHanzawa2013,OharaHanzawa2014}.
Inside the Kondo insulators, there exist the Kondo excitons which are bound states of quasielectrons and quasiholes as excited states upon the interaction-induced ground state~\cite{DuanArovasSham1997,FuhrmanNikolic2014,Nikolic2014,Fuhrmanetal2015,KnollCooper2017,ChowdhurySodemannSenthil2018}.
The Kondo excitons affect the transport phenomena, such as sound waves and heat conduction accompanied by the number and energy fluctuations, respectively, while they are usually irrelevant to electric conductivity due to the neutral charge~\cite{Sato2019}.
Thus, research of Kondo excitons is important to unveil the nonperturbative aspect in Kondo insulators, leading to understanding the nature of composite particles in heavy-fermion systems.
Recently, active researches are devoted to topological Kondo insulators, e.g., SmB$_{6}$ and YbB$_{12}$~\cite{DzeroSunGalitskiColeman2010,KuritaYamajiImada2011,DzeroSunColemanGalitski2012,TranTakimotoKim2012,AlexandrovDzeroColeman2013,WernerAssaad2013,LuZhaoWengFangDai2013} (see also Refs.~\cite{ZahidXuNeupane2015,Das2016,DzeroXiaGalitskiColeman2016,Rachel2018,ZhangZhuZhaoZhu2018} and references therein).

\begin{figure}[t!]
\centering
\includegraphics[scale=0.24]{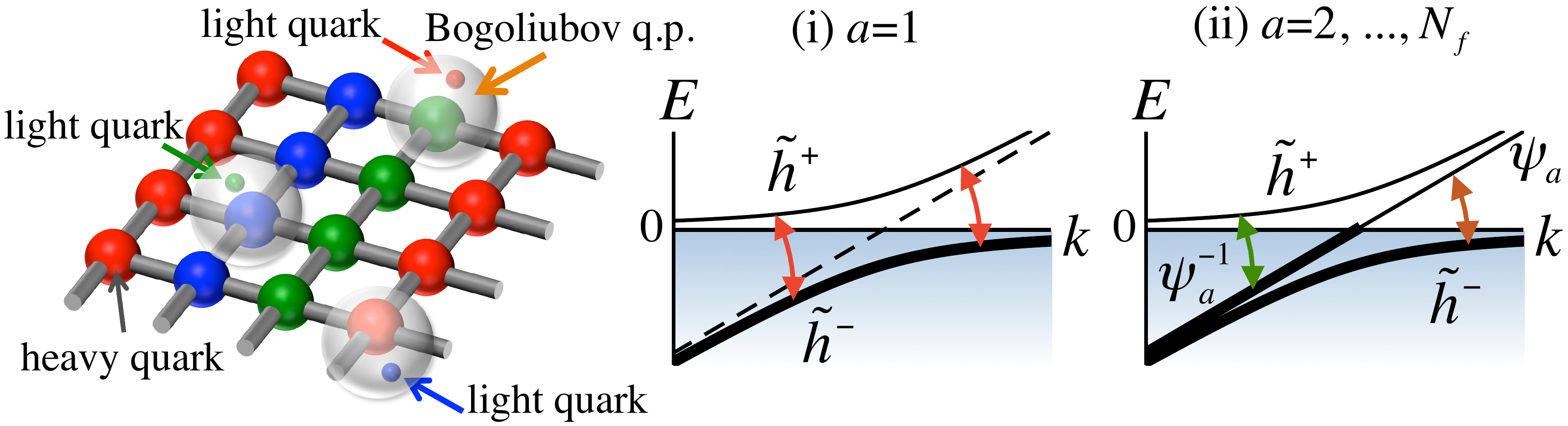}
\caption{The schematic figures of the Bogoliubov quasiparticle (q.p.) and the dispersion relations in the QCD KI ($E$, energy; $k$, momentum). The linear (solid, dotted) lines are the free quark states. The thick lines indicate the occupied states below the Fermi surface at $E\!=\!0$. The arrows in (i) and in (ii) indicate the QCD Kondo excitons.}
\label{fig:B_dispersion}
\end{figure}

Recent theoretical studies proposed that the Kondo effect arises even in matter much different from electron or atom systems, i.e., in superdense matter governed by the strong interaction described based on quantum chromodynamics (QCD), namely, nuclear matter and quark matter, which can really exist in extreme environments such as ultra-relativistic heavy-ion collisions and possibly also neutron and/or quark stars~\cite{Yasui:2013xr,Hattori:2015hka,Ozaki:2015sya,Yasui:2016ngy,Yasui:2016svc,Yasui:2016hlz,Yasui:2016yet,Kanazawa:2016ihl,Kimura:2016zyv,Yasui:2017izi,Suzuki:2017gde,Yasui:2017bey,Kimura:2018vxj,Hattori:2019zig,Yasui:2019ogk,Macias:2019vbl}.
Interestingly, the Kondo effect is accessible also in the lattice QCD simulations, as explained later in details.
We consider the situation that the heavy (charm or bottom) quarks exist as impurity particles in the quark matter, {\it i.e.}, the gas composed of up, down, and strange quarks.
In the quark matter, the gluon exchange between a light quark and a heavy quark provides the non-Abelian interaction with the color $\mathrm{SU}(N_{c})$ symmetry, where $N_{c}$ is the number of colors.
This is called {\it the QCD Kondo effect}~\cite{Yasui:2013xr,Hattori:2015hka}.
For simplicity,
we consider an ideal situation that heavy quarks are distributed statically and uniformly as impurity particles in quark matter~\cite{Yasui:2016svc,Yasui:2017izi,Kanazawa:2016ihl,Suzuki:2017gde}.
According to the mean-field calculation, a light quark and a heavy quark are mixed to form the so-called Bogoliubov quasiparticles with a gap (or mass), i.e., the condensate of the light and heavy quarks (Fig.~\ref{fig:B_dispersion}).
The unconventional phase with such a nonzero gap is called {\it the QCD Kondo phase}~\cite{Yasui:2016svc,Yasui:2017izi}.
We also regard its ground state as {\it the QCD Kondo insulator (KI)}. Here, the ground state possesses nontrivial topological charges
due to the existence of the monopoles induced by the Berry phase~\cite{Yasui:2017izi}, so that we may call this phase {\it the topological QCD KI}.

\begin{table}[b!]
\caption{Properties of the carriers in the normal phase and the QCD Kondo phase (KE: Kondo exciton).  For $N_{f}\!=\!3$, $Q_{u}\!=\!+2/3$ and $Q_{d}\!=\!Q_{s}\!=\!-1/3$. The normal phase means the non-interacting light-quark gas ($N_{f}$ flavors).
The asterisk in ``$1^{\ast}$" indicates that the light-flavor is mixed with the heavy flavor.
We show possible color representations, ${\mathbf 8}$, $\bar{\mathbf 3}$, and ${\mathbf 6}$, which are not covered in the present paper.}
\label{table:QCD_Kondo_excitons}
\begin{center}
\renewcommand{\arraystretch}{1.3}
\begin{tabular}{lcccc}
\hline
\hline
           & Normal & \multicolumn{3}{c}{QCD Kondo phase} \\
 Carrier  & $q$ & $q$ & Dressed KE & Half-dressed KE \\
\hline
  Color & ${\mathbf 3}$ & ${\mathbf 3}$ & ${\mathbf 1}$, (${\mathbf 8}$, $\bar{\mathbf 3}$, ${\mathbf 6}$) & ${\mathbf 1}$, (${\mathbf 8}$, $\bar{\mathbf 3}$, ${\mathbf 6}$)  \\
  Flavor no. & $N_{f}$ & $N_{f}\!-\!1$ & $1^{\ast}$ & $N_{f}\!-\!1$ \\
  Elec. charge & $Q_{q}$ & $Q_{q}$ & $0$ & $0$, $\pm1$ \\
  Spin (parity) & $1/2$ & $1/2$ & $0^{\pm}$, $1^{\pm}$, $\dots$ & $0^{\pm}$, $1^{\pm}$, $\dots$ \\
\hline
\hline
\end{tabular}
\renewcommand{\arraystretch}{1}
\end{center}
\end{table}%

In this paper, we discuss ``exciton" modes playing a fundamental role for transport phenomena in the QCD KI, which are bound states induced by the attraction between a Bogoliubov (quasi)particle and a (quasi)hole.
We call them {\it the QCD Kondo excitons}, or just the Kondo excitons for short.
In Table~\ref{table:QCD_Kondo_excitons}, we summarize the possible properties of the Kondo excitons. 
We emphasize that such excitons can be neutral in color (${\mathbf 1}$) and/or in electric charges ($0$), while they can also be colorful (${\mathbf 8}$, $\bar{\mathbf 3}$, $\bar{\mathbf 6}$) and/or electrically charged ($\pm1$).
If neutral excitons appear, there can exist neutral currents carrying only heat and sound waves without conducting color and electric charges. The neutral currents can be observed as violation of the so-called Wiedemann-Franz law between the electric and thermal conductivities (see e.g., Ref.~\cite{Sato2019}). Such phenomena are definitely distinguished from the conventional situation in normal light-quark gas that any current should inevitably accompany the movement of color (${\mathbf 3}$) and electric charges ($+2/3$, $-1/3$).
In the present paper, we study the mass spectrum and the energy-momentum dispersion relations of the QCD Kondo excitons, and show that there exist various types with different quantum numbers ($J^{P}$: the total spin $J$ and parity $P$).

This paper is organized as follows. In Sec.~\ref{sec:Model}, we introduce a model to describe the QCD Kondo effect. In Sec.~\ref{sec:Results}, we show the procedure to derive the appearance of Kondo excitons and the resultant dispersion relations for them. In Sec.~\ref{sec:Conclusions}, we conclude our paper and comment on discussions which are not covered in the present paper.

\section{Model}
\label{sec:Model}

We start with the Nambu--Jona-Lasinio (NJL)-type Lagrangian describing four-point interaction between the light (massless) quark and the heavy quark~\cite{Yasui:2013xr,Yasui:2016svc,Yasui:2017izi,Yasui:2016yet},
\begin{align}
\hspace{-0.3em}
{\cal L} 
&\equiv \bar{\psi}_a
\bigl(i\partial_{\mu}\gamma^{\mu}
+\mu\gamma_0\bigr)\psi_a+\Psi^{\dag}i\partial_{0}\Psi   -\lambda \bigl(\Psi^{\dag}\Psi -n_Q\bigr)  
\nonumber \\
& +G\Bigl[|\bar{\psi}_a\Psi|^2 +|\bar{\psi}_ai\gamma_5\Psi|^2 + |\bar{\psi}_a \gamma^{i}\Psi|^2  +|\bar{\psi}_a\gamma^{i}\gamma_5\Psi|^2\Bigr] ,
\label{StartL}
\end{align}
with the time-space derivatives $\partial_{\mu}$ ($\mu=0,1,2,3$) and the Dirac matrices $\gamma_{0}=\gamma^{0}$, $\gamma_{i}=-\gamma^{i}$ ($i=1,2,3$), and $\gamma_{5}$.
The symmetry is governed by the chiral symmetry for the light quark and the spin symmetry for the heavy quark.
In the first line, $\psi_a$ is the relativistic spinor fields for the light quarks with the chemical potential $\mu$ ($N_f$ flavors: $a=1,2,\cdots, N_f$), and $\bar{\psi}_{a}\equiv\psi_{a}^{\dag}\gamma_{0}$.
Notice that the repeated indices are summed over.
$\Psi$ is the nonrelativistic spinor field for the heavy quark defined in the rest frame in the context of the heavy quark effective theory (HQET)~\cite{Eichten:1989zv,Georgi:1990um} (see also Refs.~\cite{Manohar:2000dt,Neubert:1993mb,Casalbuoni:1996pg} for reviews).
According to the prescription in the HQET, the mass term of the heavy quark is omitted.
We suppose that the heavy quarks, which are regarded to be sufficiently heavy, exist as impurities with the number density  $n_Q$, and introduce $\lambda$ as a Lagrange multiplier to keep the constraint $\Psi^{\dag}\Psi=n_Q$ satisfied always~\cite{Yasui:2016svc,Yasui:2017izi,Yasui:2016yet}.
In the second line in Eq.~(\ref{StartL}), the interaction terms with the coupling constant $G$ are obtained in the leading order of the large $N_c$ expansion for the interaction deduced by a one-gluon exchange with the electric Debye mass as the leading order for the quark-gluon coupling in the QCD~\cite{Hattori:2015hka}.
We comment that the interaction strength in Eq.~\eqref{StartL} can generally be different from that estimated from other approaches, e.g., an interaction induced by instantons~\cite{Diakonov:1989un,Chernyshev:1995gj}, and others~\cite{Bardeen:1993ae,Ebert:1994tv,Mota:2006ex,Blaschke:2011yv,Guo:2012tm}.
However, such a quantitative difference does not change our conclusion (e.g., the appearance of neutral currents) qualitatively.

In order to search the ground state, we rewrite the Lagrangian (\ref{StartL}) by introducing the auxiliary fields
\begin{align}
&
\sigma_a \equiv G \, \bar{\psi}_a\Psi, \
\pi_a \equiv G \, \bar{\psi}_ai\gamma_5\Psi, \  \nonumber \\ &
V^i_a \equiv G \, \bar{\psi}_a\gamma^i\Psi, \ 
A^i_a \equiv G \, \bar{\psi}_a\gamma^i\gamma_5\Psi,
\end{align}
for the bosonic excitations composed of the light quark and the heavy quark (the heavy-light modes) as the effective degrees of freedom at low energy. 
Then, the generating functional for the Lagrangian~(\ref{StartL}) is expressed by functional integrals in terms of $\sigma_a$, $\pi_a$, $V^i_a$, and $A^i_a$ in addition to $\psi_a$ and $\Psi$.
For the ground state, we introduce the mean-field approximation in the so-called {\it hedgehog ansatz} for $a=1$: $\langle \sigma_1 \rangle = \Delta$ and $\langle V^i_1 \rangle = \Delta \, \hat{k}^i$ with the condensate (complex amplitude) $\Delta$, and the others zero~\cite{Yasui:2016svc,Yasui:2017izi}.~\footnote{This ansatz is also applicable to the case of a single heavy quark~\cite{Yasui:2016yet}.}
Here $\hat{k}^i \equiv k^i/k$ ($k\equiv|{\bm k}|$) is a unit vector along the direction of the (residual) momentum.
With this ansatz, we find that the light quark for $a=1$ and the heavy quark are superposed to form the Bogoliubov quasiparticle, and that the energy-momentum dispersions are
\begin{align}
E_k^\pm \equiv \frac{1}{2}\biggl(k-\mu+\lambda\pm\sqrt{\bigl(k-\mu-\lambda\bigr)^2+8|\Delta|^2}\biggr) , 
\label{eq:mixing_deispersion}
\end{align}
for the positive-energy states 
and $\tilde{E}_k \equiv -k-\mu$ for the negative-energy state. Notice that, in Eq.~\eqref{eq:mixing_deispersion}, flavor mixing is induced by the strong coupling in the Kondo effect.
The other light quarks for $a=2,\dots,N_{f}$ remain to be free states with the energy-momentum dispersions $E_{k}=k-\mu$ and $\tilde{E}_{k}$.
We denote $\psi_{a}$ for the particle-state and $\psi^{-1}_{a}$ for the hole state ($a=2,\dots,N_{f}$). 

Upon the above treatment, we obtain the effective action $\Gamma_{\rm eff}[\{\Phi\};\Delta]$ by performing the loop expansion at the leading order. 
Here $\{\Phi\}$ stands for $\sigma_a$, $\pi_a$, $V^i_a$, and $A^i_a$ collectively as the deviation from the mean field.
The ground state is determined self-consistently by the gap equation $\partial {\cal V}_{\rm eff}[\Delta]/\partial \Delta = 0$ stemming from the saddle-point condition for the effective potential ${\cal V}_{\rm eff}$ defined by $\Gamma_{\rm eff}[\{\Phi=0\};\Delta] \equiv -{\cal V}_{\rm eff}[\Delta]V$ with the time-space volume $V$. 
In the following analysis, we set $\lambda=0$ for simplicity, and denote $\tilde{h}^+$ for a quasiparticle with $E_k^+>0$ and $\tilde{h}^-$ a quasihole with $E_k^-<0$.
Numerically solving the gap equation, we obtain $|\Delta|=0.0845$ GeV at $\mu=0.5$ GeV by choosing the conventional parameters, $G=9.47$ GeV$^{-2}$ and the three-dimensional cutoff $\Lambda = 0.65$ GeV for the loop functions~\cite{Yasui:2016svc,Yasui:2017izi}.
In the weak-coupling limit, the condensate can be expressed approximately by the analytic form, $|\Delta| \approx \alpha \exp\bigl(-2\pi^{2}/(4N_{c}G\mu^{2})\bigr)$ with the dimensionful coefficient $\alpha$~\cite{Yasui:2016svc,Yasui:2017izi}.
We investigate the dynamics of the excitations, $\{\Phi\}$, as the quantum fluctuations around the ground state. The detailed computations to derive the dispersion relations for the Bogoliubov quasiparticle and to search for the ground state are provided in Appendix~\ref{sec:EffectiveAction}.

\section{Calculations and results}
\label{sec:Results}

First, we focus on the excitation modes between the Bogoliubov quasiparticles and holes ($\tilde{h}^{\pm}$) in $a=1$.
We call them {\it the dressed Kondo excitons}. The inverse propagator is given by the second derivatives of $\Gamma_{\rm eff}[\{\Phi\};\Delta]$ with respect to $\Phi$.
We find that there are four independent channels, $(\sigma, V^{3})$, $(\pi, A^{3})$, $(V^{1}, A^{2})$, and $(V^{2}, A^{1})$, when we choose the direction of the exciton's propagations along the third axis of the space coordinate, where we omit the indices $a$ for simplicity.
We notice that $\Phi$ and $\Phi^{\dag}$ can mix with each other.~\footnote{We remark that the channel mixings, such as $\sigma$ and $V^{3}$ in $(\sigma,V^{3})$. These anomalous effects reduce the masses of both the dressed and half-dressed Kondo excitons, in which the latter will be shown later.} We find that the mixings in $(V^{1}, A^{2})$ and $(V^{2}, A^{1})$ are induced by the term proportional to $\epsilon^{ijk}\partial^iV^{j} A^{k}$ ($i,j,k=1,2,3$).
The inverse propagator for $(\sigma, V^{3})$, for instance, takes the $4 \times 4$ matrix in momentum space
\begin{align}
{\cal D}_{\sigma V^3}^{-1}= \left(
\begin{array}{cccc}
\bar{D}_{\sigma^\dagger\sigma} & \bar{D}_{\sigma^\dagger V^3} & \bar{D}_{\sigma^\dagger \sigma^\dagger}  & \bar{D}_{\sigma^\dagger V^{3\dagger}} \\
\bar{D}_{V^{3\dagger}\sigma} & \bar{D}_{V^{3\dagger} V^3} &\bar{D}_{V^{3\dagger} \sigma^\dagger} & \bar{D}_{V^{3\dagger} V^{3\dagger}} \\
\bar{D}_{\sigma\sigma}  & \bar{D}_{\sigma V^3}  & \bar{D}_{\sigma \sigma^\dagger} & \bar{D}_{\sigma V^{3\dagger}} \\
\bar{D}_{V^{3}\sigma} & \bar{D}_{V^{3} V^3}  & \bar{D}_{V^{3} \sigma^\dagger}  & \bar{D}_{V^{3} V^{3\dagger}} \\
\end{array}
\right) , \label{DMatrix}
\end{align}
in which the matrix element $\bar{D}_{\sigma^\dagger\sigma}(q)$ with four-momentum $q=(q_{0},{\bm q})$ is defined by
\begin{align}
\frac{\delta^2\Gamma_{\rm eff}[\{\Phi\};\Delta]}{\delta \tilde{\sigma}^\dagger(p) \, \delta\tilde{\sigma}(q)} = \bar{D}_{\sigma^\dagger\sigma}(q)(2\pi)^4\delta^4(p+q) ,  \label{TwoPointMS}
\end{align}
where $\tilde{\sigma}(p)$ with four-momentum $p$ is defined by the Fourier transformation of $\sigma$,
and the other elements are also defined. The inverse propagators ${\cal D}_{\pi A^3}^{-1}$, ${\cal D}_{V^1 A^2}^{-1}$, and ${\cal D}_{V^2 A^1}^{-1}$ are obtained just as in Eq.~\eqref{DMatrix}. The detail of the matrix elements of the inverse propagator is shown in Appendix~\ref{sec:Excitons}.
The dispersions for the dressed Kondo excitons are given as the solutions of ${\rm det}\bigl({\cal D}_{\sigma V^3}^{-1}\bigr)=0$, ${\rm det}\bigl({\cal D}_{\pi A^3}^{-1}\bigr)=0$, ${\rm det}\bigl({\cal D}_{V^1 A^2}^{-1}\bigr)=0$, and ${\rm det}\bigl({\cal D}_{V^2 A^1}^{-1}\bigr)=0$, respectively, in each channel.

\begin{figure}[tb]
\centering
\includegraphics[scale=0.3]{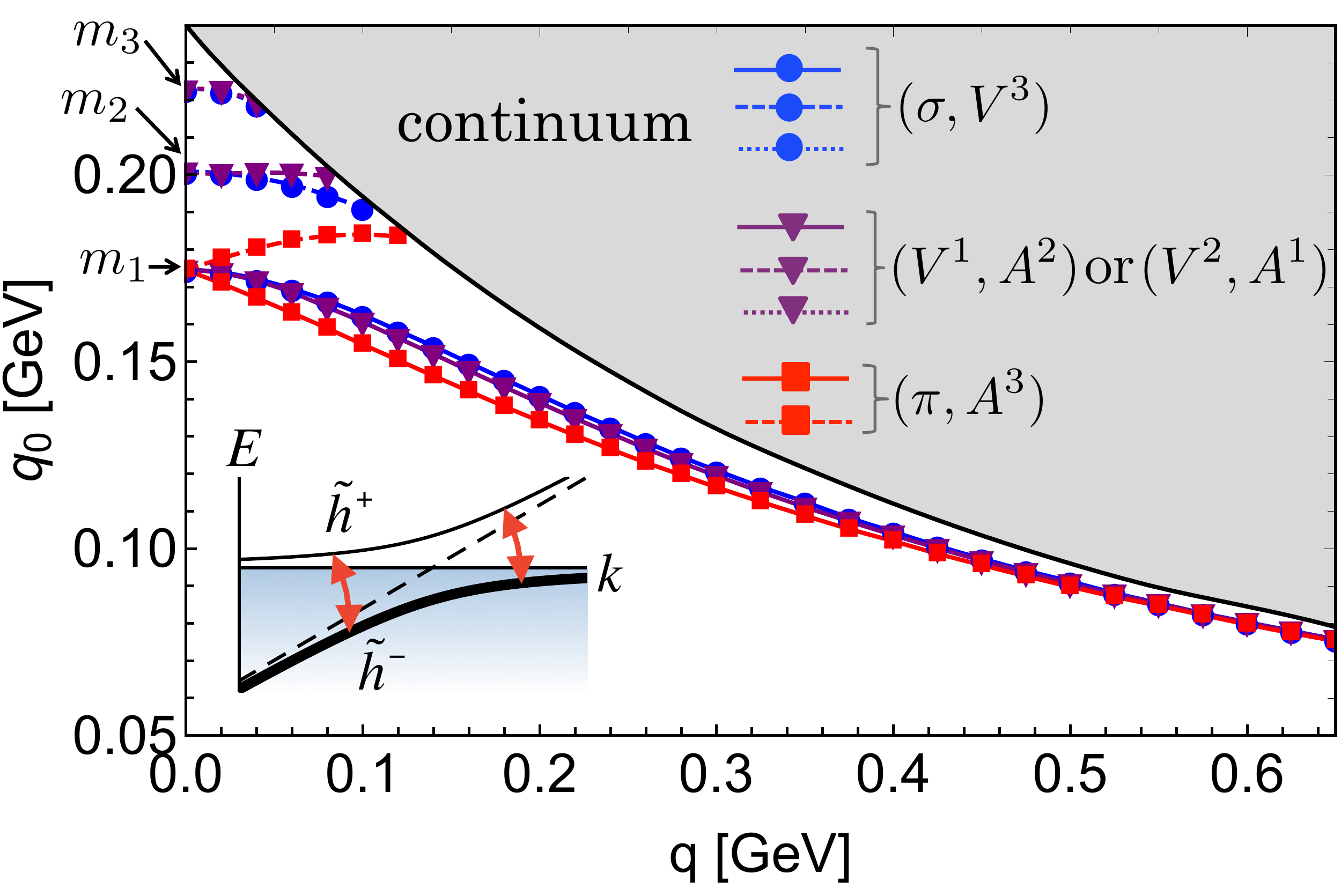}
\caption{The dispersion relations for the dressed Kondo excitons, $(q_{0},q)$ with $q=|{\bm q}|$, in light flavor $a=1$. See also Fig.~\ref{fig:B_dispersion}(i).}
\label{fig:PlotNf1}
\end{figure}

\begin{figure}[tb]
\centering
\includegraphics*[scale=0.3]{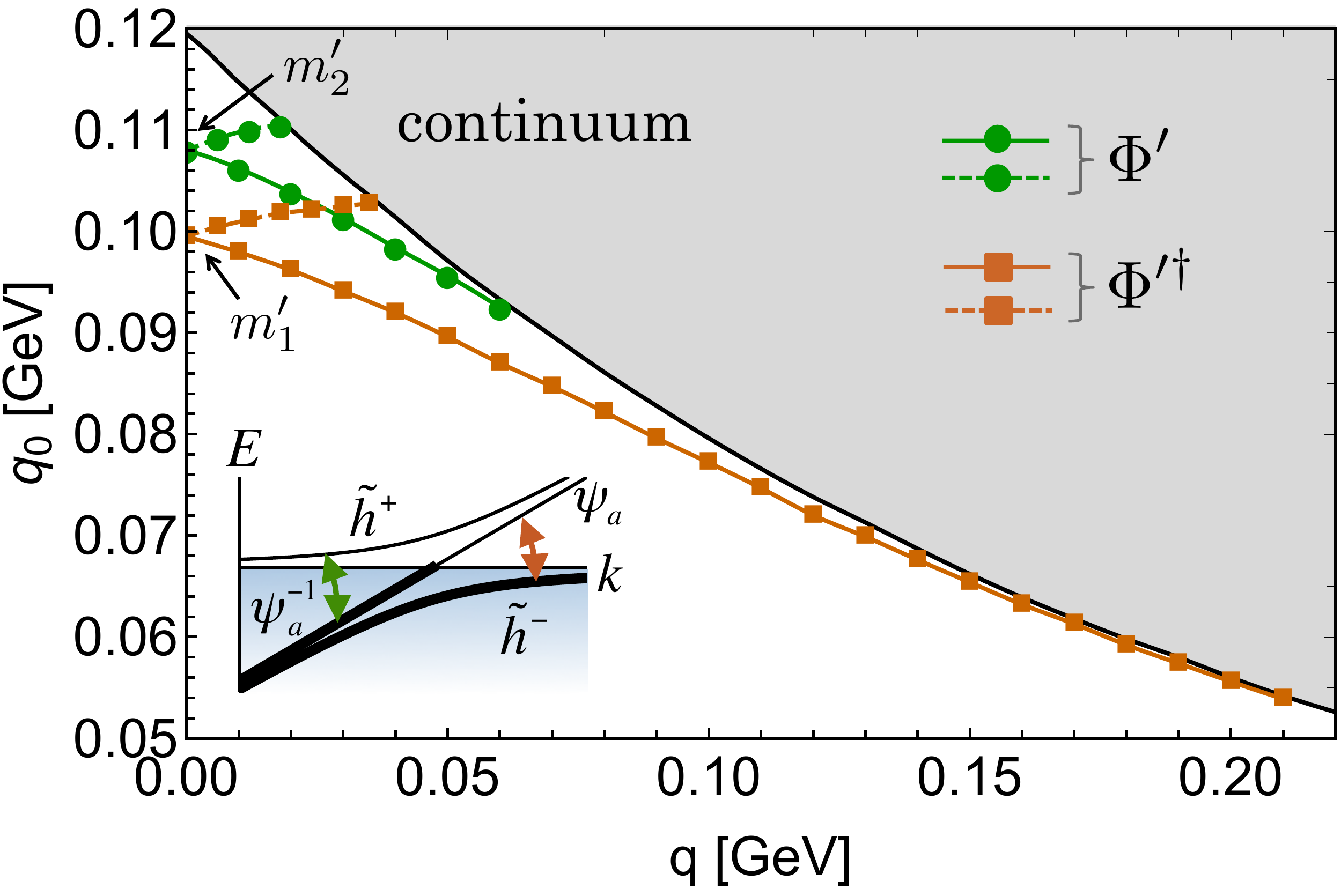}
\caption{The dispersion relations for the half-dressed Kondo excitons, $(q_{0},q)$ with $q = |{\bm q}|$, in light flavor $a=2,\dots,N_{f}$. ${\Phi'}$ denotes $(\sigma,V^3)$, $(\pi,A^3)$, $(V^1, A^2)$, and $(V^2,A^1)$ collectively. See also Fig.~\ref{fig:B_dispersion}(ii).}
\label{fig:PlotNf2}
\end{figure}

We show the numerical results of the dispersion relations $(q_{0},q)$ with $q=|{\bm q}|$ in Fig.~\ref{fig:PlotNf1}.
The gray region indicates a continuum regime for a pair of $\tilde{h}^+$ and $\tilde{h}^-$.
Under the continuum region, there are the dressed Kondo excitons in $(\sigma, V^3)$ and $(\pi, A^3)$, as shown by the blue and red curves, respectively.
The dispersion relations in $(V^1,A^2)$ and $(V^2, A^1)$, which are shown by the purple curves, coincide with each other due to the rotational symmetry on the transverse plane perpendicular to the third axis. 
Those Kondo excitons are the bound states of a pair of $\tilde{h}^+$ and $\tilde{h}^-$ below the thresholds, and they have no open channel for decays.
At the vanishing momentum ($|{\bm q}| = 0)$, there exist three dressed Kondo excitons with masses $m_1= 0.175$  GeV (quintuply degenerate), $m_2=0.201$ GeV (triply degenerate), and $m_3 = 0.223$ GeV (triply degenerate).
In this limit, the channel mixings such as between $\sigma$ and $V^{3}$ are independently resolved: the $m_1$ state is given by either of $\sigma$, $\pi$, or $A^{i}$, while the $m_2$ and $m_3$ states are given by any of $V^{i}$ ($i=1,2,3$).

Second, we consider the exciton modes between $\tilde{h}^{+}$ and $\psi^{-1}_{a}$ as well as between $\tilde{h}^{-}$ and $\psi_{a}$
with the light flavor $a=2,\dots,N_{f}$.
We call them {\it the half-dressed Kondo excitons}, because one of the fermions composing the exciton is a free quark.
In this case, the boson number is conserved, and hence the inverse-propagator alternative to Eq.~\eqref{DMatrix} takes the form
\begin{eqnarray}
{\cal D}_{\sigma V^3}^{\prime\,-1}= \left(
\begin{array}{cccc}
\bar{D}^{\prime}_{\sigma^\dagger\sigma} & \bar{D}^{\prime}_{\sigma^\dagger V^3} & 0 & 0 \\
\bar{D}^{\prime}_{V^{3\dagger}\sigma} & \bar{D}^{\prime}_{V^{3\dagger} V^3} & 0 & 0 \\
0&0& \bar{D}^{\prime}_{\sigma \sigma^\dagger} & \bar{D}^{\prime}_{\sigma V^{3\dagger}} \\
0 & 0  & \bar{D}^{\prime}_{V^{3} \sigma^\dagger}  & \bar{D}^{\prime}_{V^{3} V^{3\dagger}} \\
\end{array}
\right) , \label{DMatrix2}
\end{eqnarray}
with the matrix elements defined similarly to Eq.~\eqref{TwoPointMS},
and so are the other inverse propagators ${\cal D}^{\prime\,-1}_{\pi A^3}$, ${\cal D}^{\prime\,-1}_{V^1A^2}$, and ${\cal D}^{\prime\,-1}_{V^2A^1}$, where the indices $a=2,\dots,N_{f}$ in $\sigma$, $\pi$, $V^{i}$, and $A^{i}$ are omitted.
The conservation of the light flavor ($a$) and the heavy-quark spin yields the simple relations: ${\cal D}_{\sigma V^3}^{\prime\,-1} = {\cal D}^{\prime\,-1}_{\pi A^3} = {\cal D}^{\prime\,-1}_{V^1A^2} = {\cal D}^{\prime\,-1}_{V^2A^1}$.
Thus, we find that the matrix in Eq.~\eqref{DMatrix2} is separated into the two sectors represented by the top-left submatrix and the bottom-right submatrix.
Each sector includes the propagation of the excitation of $\tilde{h}^{+}\psi^{-1}_{a}$ (${\Phi'}$) or $\tilde{h}^{-}\psi_{a}$ (${\Phi'}^\dagger$) for $a=2,\dots,N_{f}$, where ${\Phi'}$ denotes $(\sigma,V^3)$, $(\pi,A^3)$, $(V^1, A^2)$, and $(V^2,A^1)$ collectively. As is the case with $a=1$, the detail of the matrix elements of the inverse propagator is shown in Appendix~\ref{sec:Excitons}.

We show the numerical results of the dispersion relations for the half-dressed Kondo excitons in Fig.~\ref{fig:PlotNf2}. The green (orange) curves correspond to ${\Phi'}$ (${\Phi'}^\dagger$).
The gray region indicates the continuum regime for a pair of $\tilde{h}^+$ and $\psi^{-1}_{a}$ in ${\Phi'}$ as well as for a pair of $\tilde{h}^-$ and $\psi_{a}$ in ${\Phi'}^\dagger$.
At the vanishing momenta, we find the two types of half-dressed Kondo excitons with the masses $m'_{1}=0.100$ GeV (eight-fold degenerate) and $m'_{2}=0.108$ GeV (eight-fold degenerate), respectively.
The mass difference between the two modes stems from the violation of the particle-hole symmetry.

We observe some differences between the dressed Kondo excitons ($a=1$) and the half-dressed Kondo excitons ($a=2,\dots,N_{f}$).
First, there are many branches of the dispersion relations in the dressed Kondo excitons, while most of them are degenerate in the half-dressed Kondo excitons.
This difference stems from the nonconservation or the conservation of the boson number.
Second, there is a tendency that the excitation energies of the dressed Kondo excitons are higher than those of the half-dressed Kondo excitons.
This result indicates that, in $N_{f}\!\ge\!2$, the dressed Kondo excitons are the higher-excited states, while the half-dressed Kondo excitons are the lower-excited states.

The Kondo excitons considered in this paper are neutral in color, and hence there is no color current at low energy.
As for the electric charge, there are subtle differences between the dressed and half-dressed Kondo excitons.
The dressed Kondo excitons are neutral in electric charge, while the half-dressed Kondo excitons are either neutral or charged.
For example, when the electric charges of $\tilde{h}^{+}$ and $\psi_{a}^{-1}$ (or $\tilde{h}^{-}$ and $\psi_{a}$) are different, the exciton is electrically charged (Table~\ref{table:QCD_Kondo_excitons}).

\section{Conclusions and discussions}
\label{sec:Conclusions}

We have shown the properties of the QCD Kondo excitons as the excited states in the QCD KI, and that they can induce neutral currents for color and/or electric charge.
This will play a crucial role in the transport phenomena in the QCD KI.

Some discussions are ready for the present treatment of the QCD KI.
First, in the construction of the Bogoliubov quasiparticle, we assumed that the electric charges of the light quark and the heavy quark are the same.
If the electric charges are different from each other, we would encounter a problem of the spontaneous breaking of the $\mathrm{U}(1)$ gauge symmetry for electric charges, i.e., the Higgs mechanism.
We have to reformulate the procedure of analysis by considering the gauge field as a dynamical degree of freedom. 
Second, we have ignored the color superconductivity induced by the attraction between two light quarks~\cite{Alford:1997zt,Rapp:1997zu,Alford:1998mk} (see also Refs.~\cite{Rajagopal:2000wf,Alford:2007xm,Fukushima:2010bq,Fukushima:2013rx} for reviews).
Though the superconductivity gap is suppressed in the large $N_{c}$~\cite{Shuster:1999tn}, we will need to carefully examine how the color superconductor at ${\cal O}(1/N_{c})$ would affect the QCD Kondo excitons.
Note that the condensate and the excitons in the QCD KI are color neutral, and hence they survive in the large $N_{c}$ limit.
Third, we have not considered the surface effect, i.e., zero-mode fermions at the surface of the topological QCD KI.
Such zero modes can contribute to the transport phenomena.

We mention the applicability of numerical simulations in the lattice QCD formalism to the QCD KI in special settings.
Notice that the condensate and the excitons in the QCD KI are color neutral, so that they are gauge-invariant observables.
We consider the $u\bar{d}$ quark matter at {\it zero} baryon chemical potential and {\it nonzero} isospin chemical potential.
This system can be simulated by avoiding the difficulty of the so-called sign problem~\cite{Kogut:2002tm,Kogut:2002zg,Kogut:2004zg,Brandt:2017oyy}.
When the heavy quarks exist as impurities, there will be the dressed Kondo excitons composed of $u$ quark holes and heavy quarks, or $\bar{d}$ quarks and heavy quarks.~\footnote{Notice that the pion condensate is competitive to the QCD Kondo effect.}
The correlation function of color-neutral currents, in which the contribution from the dressed Kondo excitons should be relevant,
is the gauge-invariant quantities measurable in lattice QCD simulations.
The signal of the dressed Kondo excitons can be distinguished from that of the confining heavy-light ($D$ or $B$) mesons, since the quark matter at a large isospin chemical potential should be the deconfined phase.

In recent studies, new types of the QCD Kondo effect were proposed: the QCD Kondo effect in strong magnetic fields~\cite{Ozaki:2015sya} and the QCD Kondo effect induced by the gapped quark in the two-flavor color superconductor~\cite{Hattori:2019zig}.
They suggest that the QCD Kondo effect can emerge in various environments relating to the strong interaction.
There are also many other topics: the Kondo resonance for a single heavy quark~\cite{Yasui:2016yet}, the competitions between the QCD Kondo phase and the chiral condensate~\cite{Suzuki:2017gde} or the color superconductor~\cite{Kanazawa:2016ihl}, the conformal theory for the Fermi and non-Fermi liquids~\cite{Kimura:2016zyv,Kimura:2018vxj}, and the transport properties~\cite{Yasui:2017bey}.
For more advanced studies, we may consider the continuity from the Kondo effect in the deconfinement phase to the Kondo effect for heavy hadrons in the confinement phase (nuclear matter)~\cite{Yasui:2013xr,Yasui:2016ngy,Yasui:2016hlz,Yasui:2019ogk} (see, e.g., Refs.~\cite{Hosaka:2016ypm,Krein:2017usp} for review for heavy hadrons in nuclear matter).
We may raise a question whether the Kondo phase in the quark matter is continuously connected to that in the hadronic matter~\cite{Hattori:2015hka}.
Applications of the QCD Kondo effect to charm stars are also interesting~\cite{Macias:2019vbl}.
Studying those problems in terms of the Kondo excitons should be left for future works.

\section*{Acknowledgement}

The authors thank Muneto Nitta, Yoshimasa Hidaka, Yasufumi Araki, and Daisuke Inotani for fruitful discussions. This work is supported by National Natural Science Foundation of China Grant No.~20201191997 (D.~S.), by a Japan Society for the Promotion of Science Grant-in-Aid for Scientific Research [KAKENHI Grants No.~JP17K14277 (K.~S.) and No.~17K05435 (S.~Y.)], and by a Ministry of Education, Culture, Sports, Science, and Technology supported Program for the Strategic Research Foundation at Private Universities ``Topological Science" (Grant No. S1511006) (S.~Y.).


\appendix

\section{Effective action}
\label{sec:EffectiveAction}

To investigate the Kondo excitons in quark matter, we start the discussion with the NJL-type Lagrangian describing the four-point interaction between a light and a heavy quark.
This interaction was first applied to the Kondo effect without nonperturbative contribution in Ref.~\cite{Yasui:2013xr}, and then it was utilized as a mean-field approach to study the nonperturbative QCD Kondo effect~\cite{Yasui:2016svc,Yasui:2017izi,Yasui:2016yet}.
The Lagrangian is given by
\begin{align}
\hspace{-0.3em}
{\cal L} 
&\equiv \bar{\psi}_a
\bigl(i\Slash{\partial} +\mu\gamma_0\bigr)\psi_a+\Psi^{\dag}i\partial_{0}\Psi   -\lambda \bigl(\Psi^{\dag}\Psi -n_Q\bigr)  
\nonumber \\
& +G\Bigl[|\bar{\psi}_a\Psi|^2 +|\bar{\psi}_ai\gamma_5\Psi|^2 + |\bar{\psi}_a \gamma^{i}\Psi|^2  +|\bar{\psi}_a\gamma^{i}\gamma_5\Psi|^2\Bigr] \ ,
\label{NJL}
\end{align}
with $\Slash{\partial} \equiv \gamma^\mu\partial_\mu$ and $\bar{\psi}_a\equiv \psi_a^\dagger \gamma^0$.
The summation is implicitly taken when the indices are repeated.
In this Lagrangian, $\psi_a$ ($a=1,2,\cdots, N_f$) and $\Psi$ are the light quark fields with $N_f$ flavors and the heavy quark field carrying only its particle degree of freedom.
$\psi_a$ is a four-component spinor, and $\Psi$ is a two-component one. $\mu$ is a light quark chemical potential and $\lambda$ is a Lagrange multiplier for the condition $\Psi^\dagger\Psi=n_Q$ ($n_Q$, a space-averaged number density of heavy quarks) being always satisfied~\cite{Yasui:2016svc,Yasui:2017izi}.

 To describe bosonic degrees of freedom, we introduce auxiliary fields
\begin{align}
&
\sigma_a \equiv G \, \bar{\psi}_a\Psi, \
\pi_a \equiv G \, \bar{\psi}_ai\gamma_5\Psi, \  \nonumber \\ &
V^i_a \equiv G \, \bar{\psi}_a\gamma^i\Psi, \ 
A^i_a \equiv G \, \bar{\psi}_a\gamma^i\gamma_5\Psi, \label{Auxiliary}
\end{align}
for the heavy-light modes,
in which $\sigma_a$, $\pi_a$, $V_a^i$, and $A_a^i$ are the scalar, pseudoscalar, vector, and axialvector, respectively. For the ground state, we assume the so-called hedgehog ansatz in the momentum space:
\begin{eqnarray}
\langle \sigma_1 \rangle = \Delta\ , \ \ \langle V_1^i \rangle = \Delta\hat{k}^i\ ,
\end{eqnarray}
where $\hat{k}^i \equiv k^i/k$ ($k= |{\bm k}|$) with ${\bm k}$ the three-dimensional momentum for $a=1$, and $\langle \sigma_a\rangle = \langle V_n^i\rangle=0$ for $a=2,3,\cdots N_f$~\cite{Yasui:2016svc,Yasui:2017izi}.
$\Delta$ is a constant mean field the value of which should be obtained by solving the gap equation.
Thus, after performing the Hubbard-Stratonovich transformation (bosonization) in terms of the auxiliary fields~\eqref{Auxiliary}, the Lagrangian~\eqref{NJL} turns into
\begin{align}
{\cal L}  &= \Big( \bar{\psi}_1\ \ \bar{\psi}_2 \ \ \cdots \ \ \bar{\psi}_{N_f} \ \ {\Psi}^\dagger \Big) \Big( {\cal G}^{-1} + {\cal V} \Big) \left(
 \begin{array}{c}
\psi_1 \\
\psi_2 \\
\vdots \\
\psi_{N_f} \\
\Psi \\
\end{array}
\right) \nonumber\\
& -\frac{1}{G}\Big(\Delta^*(\sigma_1-i\hat{\bm \nabla}\!\cdot\!{\bm V}_1) + \Delta(\sigma_1^\dagger+i\hat{\bm \nabla}\!\cdot\!{\bm V}_1^\dagger) \Big) \nonumber\\
&- \frac{1}{G}\Big(|\sigma_a|^2+|\pi_a|^2 +|V_a^i|^2 + |A_a^i|^2\Big)-\frac{2}{G}|\Delta|^2 + \lambda n_Q \ , \label{LExpand}
\end{align}
($-i\hat{\bm \nabla}\equiv -i{\bm \nabla}/|-i{\bm \nabla}|$ with ${\bm \nabla}$ being the derivative in the three-dimensional real space), where ${\cal G}^{-1}$ is an inverse-propagator matrix for a quark expressed in the flavor space
\begin{align}
{\cal G}^{-1} \equiv \left(
\begin{array}{ccccc}
{\cal G}^{-1}_{11} & 0 & \cdots  & 0 &{\cal G}^{-1}_{1h} \\
0 & {\cal G}_{22}^{-1} & \cdots & 0 & 0 \\
\vdots & \vdots & \ddots & \vdots & \vdots \\
0 & 0 & \cdots & {\cal G}^{-1}_{N_fN_f} & 0 \\
{\cal G}^{-1}_{h1}& 0 & \cdots & 0  & {\cal G}^{-1}_{hh} \\
\end{array}
\right)\ , \label{GInverse}
\end{align}
with
\begin{align}
{\cal G}^{-1}_{aa} \equiv \left(
 \begin{array}{cc}
 i\partial_0+\mu & i{\bm \nabla}\!\cdot\!{\bm \sigma} \\
- i{\bm \nabla}\!\cdot\!{\bm \sigma} & -i\partial_0-\mu\\
  \end{array}
 \right)\ \  (a=1,2,\cdots, N_f)\ , 
\end{align}
\begin{align}
{\cal G}^{-1}_{1h} \equiv \left(
 \begin{array}{c}
\Delta^* \\
\Delta^* i\hat{\bm \nabla}\!\cdot\!{\bm \sigma} \\
  \end{array}
 \right)\ , 
\end{align}
\begin{align}
{\cal G}^{-1}_{h1} \equiv \left(
 \begin{array}{cc}
\Delta & -\Delta i\hat{\bm \nabla}\!\cdot\!{\bm \sigma} \\
  \end{array}
 \right)\ , 
\end{align}
\begin{align}
{\cal G}^{-1}_{hh} \equiv (i\partial_0-\lambda ){\bm 1}\ ,
\end{align}
(${\bm 1}$ is a $2\times2$ unit matrix),
and ${\cal V}$ denotes the terms for the fluctuations of the heavy-light modes ($\sigma_a$, $\pi_a$, $V_a^i$ and $A_a^i$)
\begin{align}
{\cal V} = \left(
 \begin{array}{ccccccccc}
 0 &  0 & \cdots & 0 &{\cal V}_{1h}\\
 0 &  0 & \cdots & 0 &{\cal V}_{2h}\\
\vdots &  \vdots & \ddots & \vdots & \vdots \\
0 &  0 & \cdots & 0 &{\cal V}_{N_fh}\\
{\cal V}_{h1} &{\cal V}_{h2}& \cdots & {\cal V}_{hN_f} & 0 \\
 \end{array}
 \right)\ , \label{VMatrix}
\end{align}
with
\begin{align}
{\cal V}_{a h} =  \left(
\begin{array}{c}
\sigma_a^\dagger+{\bm A}_a^\dagger\!\cdot\!{\bm \sigma}  \\
i\pi_a^\dagger -{\bm V}_a^\dagger\!\cdot\!{\bm \sigma}\\
\end{array}
\right) \ ,
\end{align}
\begin{align}
{\cal V}_{ha} =  \left(
\begin{array}{cc}
\sigma_a + {\bm A}_a\!\cdot\!{\bm \sigma} & i\pi_a + {\bm V}_a\!\cdot\!{\bm \sigma} 
\end{array}
\right) \ ,
\end{align}
for $a=1,2,\cdots, N_f$. $\sigma^i$ ($i=1,2,3$) is the Pauli matrix.
Since $\psi_a$ is a four-component spinor and $\Psi$ is a two-component one, as mentioned before, we notice that ${\cal G}^{-1}$ and ${\cal V}$ are $(4N_f+2) \times (4N_f+2)$ dimensional matrices. In obtaining Eq.~\eqref{LExpand}, we have employed the Dirac representation for the gamma matrices:
\begin{align}
\gamma_0 = \left(
\begin{array}{cc}
1 & 0 \\
0 & -1 \\
\end{array}
\right)\ ,  \ \gamma^i =  \left(
\begin{array}{cc}
0 & \sigma^i \\
-\sigma^i & 0 \\
\end{array}
\right)\ ,  \ \gamma_5 = \left(
\begin{array}{cc}
0 & 1 \\
1 & 0 \\
\end{array}
\right)\ . \nonumber\\
\end{align}
By diagonalizing the fermion part in the Lagrangian~\eqref{LExpand}, we can show that the free fermions (without the mean field) and the Bogoliubov quasifermions with the mean field appear.
Their energy-momentum dispersion relations are obtained by solving the equations, ${\rm det}(\tilde{\cal G}^{-1}) = 0$, where $\tilde{\cal G}^{-1}$ is the Fourier transformation of ${\cal G}^{-1}$ in momentum space.
The results are
\begin{align}
E_k^\pm &\equiv \frac{1}{2}\biggl(k-\mu+\lambda\pm\sqrt{\bigl(k-\mu-\lambda\bigr)^2+8|\Delta|^2}\biggr)\ ,  \nonumber\\
\tilde{E}_k &\equiv -k-\mu \ ,
\label{EkPM}
\end{align}
for $a=1$, and
\begin{align}
E_k &\equiv k-\mu \ , \nonumber\\
\tilde{E}_k &= -k-\mu\ ,
\end{align}
for $a=2,3,\cdots, N_f$.
The results for $a=2,3,\cdots, N_f$ are the same as the free ones since $\psi_a$ ($a=2,3,.\cdots, N_f$) is not affected by the mean field at the leading approximation. In the following analysis, in order to examine the appearance of the QCD Kondo excitons in a transparent way, we will take $\lambda=0$.

For the Lagrangian~\eqref{LExpand},
we obtain an effective action $\Gamma_{\rm eff}[\{\Phi\};\Delta]$ at the one-loops of the quarks, where $\Phi$ denotes the $\sigma_a$, $\pi_a$, $V_a^i$, and $A_a^i$ collectively, which yields the following form,
\begin{align}
& \Gamma_{\rm eff}[\{\Phi\};\Delta]  \nonumber\\
&= -i{\rm Tr}\,{\rm ln}({\cal G}^{-1} +{\cal V})\nonumber\\
&-\frac{1}{G} \int d^4x\Bigg[\Delta^*(\sigma_1-i\hat{\bm \nabla}\!\cdot\!{\bm V}_1) + \Delta(\sigma_1^\dagger+i\hat{\bm \nabla}\!\cdot\!{\bm V}_1^\dagger) \nonumber\\
&\hspace{4em}
+|\sigma_a|^2+|\pi_a|^2 + |V_a^i|^2+ |A_a^i|^2 + 2|\Delta|^2 \Bigg] , \label{EffAction}
\end{align}
where ``${\rm Tr}$" represents the trace over the space-time coordinate, flavors, and Dirac indices. At the mean-field level, we find that Eq.~\eqref{EffAction} becomes
\begin{align}
\Gamma_{\rm eff}[\{\Phi=0\};\Delta]  &= -i{\rm Tr}\,{\rm ln}({\cal G}^{-1})-\frac{2V}{G}|\Delta|^2 \ ,
\end{align}
with the space-time volume $V$.
The gap equation to determine the ground state can be obtained by $\partial V_{\rm eff}/\partial\Delta= 0$, where $V_{\rm eff}$ is an effective potential defined by $V_{\rm eff} \equiv -\Gamma_{\rm eff}[\{\Phi=0\};\Delta]/V$.
As a result, the gap equation is
\begin{eqnarray}
2N_c\int^\Lambda\frac{d^3{\bm k}}{(2\pi)^3}\frac{1}{\sqrt{(k-\mu)^2+8|\Delta|^2}}-\frac{1}{G} = 0 \ , \label{GapEq}
\end{eqnarray}
in the momentum space,
where a three-dimensional cutoff $\Lambda$ is introduced for regularizing an ultraviolet divergence ($k \in [0,\Lambda]$). In obtaining Eq.~\eqref{GapEq}, we have included an infinitesimal imaginary part in the quark propagator appropriately to incorporate the existence of the Fermi surface as shown later. When we choose $\mu=0.5$ GeV, $G=9.47$ GeV$^{-2}$, and $\Lambda=0.65$ GeV, the magnitude of the gap turns out to be $|\Delta| = 0.0845$ GeV by the gap equation~\eqref{GapEq}~\cite{Yasui:2016svc,Yasui:2017izi}.

\section{Kondo Excitons}
\label{sec:Excitons}

To investigate the Kondo excitons, we need to take a second derivative with respect to $\Phi$. For this purpose, let us expand the effective action~\eqref{EffAction} up to the quadratic terms of $\tilde{\Phi}$ in the momentum space, where $\tilde{\Phi}$ is the Fourier transformation of $\Phi$. By making use of 
\begin{eqnarray}
{\rm ln}({\cal G}^{-1}+{\cal V}) = {\rm ln}({\cal G})^{-1} + {\cal G}{\cal V}- \frac{1}{2}{\cal G}{\cal V}{\cal G}{\cal V} + \cdots \ ,
\end{eqnarray}
and converting this relation into the momentum space, we obtain the effective action in terms of $\tilde{\Phi}$
\begin{align}
\Gamma_{\rm eff}[\{\tilde{\Phi}\};\Delta]  &= \Gamma_{\rm eff}[\{\tilde{\Phi}=0\};\Delta] \nonumber\\
&-\frac{i}{2} {\rm tr}\int\frac{d^4k}{(2\pi)^4}\frac{d^4q}{(2\pi)^4}\tilde{\cal G}(k)\tilde{\cal V}(-q)\tilde{\cal G}(q+k)\tilde{\cal V}(q) \nonumber\\
&-\frac{1}{G} \int \frac{d^4q}{(2\pi)^4}\Big\{|\tilde{\sigma}_a(q)|^2+|\tilde{\pi}_a(q)|^2 \nonumber\\
&+|\tilde{V}_a^i(q)|^2 + |\tilde{A}_a^i(q)|^2\Big\}  + \cdots\ , \label{GammaEx}
\end{align}
in which the gap equation~\eqref{GapEq} has been utilized, and we have defined $|\tilde{X}_a(q)|^2 \equiv \tilde{X}_a(-q)^\dagger \tilde{X}_a(q)$. The symbol ``{\rm tr}" in Eq.~\eqref{GammaEx} stands for a trace operator over the flavor and Dirac indices. $\tilde{\cal G}(k)$ is a propagator matrix for the (quasi)fermions in the momentum space.
This is obtained by the inverse matrix of the Fourier transformation of Eq.~\eqref{GInverse}, and takes the form of
\begin{align}
\tilde{\cal G} = \left(
\begin{array}{ccccc}
\tilde{\cal G}_{11} & 0 & \cdots  & 0 &\tilde{\cal G}_{1h} \\
0 & \tilde{\cal G}_{22} & \cdots & 0 & 0 \\
\vdots & \vdots & \ddots & \vdots & \vdots \\
0 & 0 & \cdots & \tilde{\cal G}_{N_fN_f} & 0 \\
\tilde{\cal G}_{h1}& 0 & \cdots & 0  & \tilde{\cal G}_{hh} \\
\end{array}
\right)\ ,
\end{align}
where the matrix elements are given by
\begin{widetext}
\begin{align}
\tilde{\cal G}_{11}(k) \equiv \frac{1}{(k_0-\tilde{E}_k-i\eta)(k_0-E_k^++i\eta)(k_0-E_k^--i\eta)} \left(
\begin{array}{cc}
(k_0+\mu)k_0-|\Delta|^2 & -\left[kk_0+|\Delta|^2\right]\hat{\bm k}\!\cdot\!{\bm \sigma} \\
\left[kk_0+|\Delta|^2\right]\hat{\bm k}\!\cdot\!{\bm \sigma}  & -(k_0+\mu)k_0+|\Delta|^2 \\\end{array}
\right) \ ,
\end{align}
\begin{align}
\tilde{\cal G}_{aa}(k) \equiv \frac{1}{k_0-\tilde{E}_k^+-i\eta}\left(\frac{\theta(\mu-k)}{k_0-E_k-i\eta}+\frac{\theta(k-\mu)}{k_0-E_k+i\eta}\right) \left(
\begin{array}{cc}
k_0+\mu  & -{\bm k}\!\cdot\!{\bm \sigma}  \\
 {\bm k}\!\cdot\!{\bm \sigma} & -k_0-\mu  \\ 
 \end{array}
\right)\ \  (a=2,3,\cdots, N_f) \ ,
\end{align}
\begin{align}
\tilde{\cal G}_{1h}(k) \equiv \frac{1}{(k_0-E_k^++i\eta)(k_0-E_k^--i\eta)} \left(
\begin{array}{c}
-\Delta^* \\
 -\Delta^*  \hat{\bm k}\!\cdot\!{\bm \sigma}
 \end{array}
\right) \ ,
\end{align}
\begin{align}
\tilde{\cal G}_{h1}(k) \equiv \frac{1}{(k_0-E_k^++i\eta)(k_0-E_k^--i\eta)} \left(
\begin{array}{cc}
 -\Delta  &\Delta  \hat{\bm k}\!\cdot\!{\bm \sigma} 
 \end{array}
\right) \ ,
\end{align}
\begin{align}
\tilde{\cal G}_{hh}(k) \equiv \frac{(k_0-k+\mu)}{(k_0-E_k^++i\eta)(k_0-E_k^--i\eta)} {\bm 1}\ ,
\end{align}
\end{widetext}
with $\hat{\bm k}={\bm k}/|{\bm k}|$.
In the above expressions, we have inserted an infinitesimal imaginary part $+i\eta$ or $-i\eta$ ($\eta>0$), when the pole of each dispersion lies above or below the Fermi surface, respectively, to incorporate the existence of the Fermi surface appropriately.

Inverse propagators for the fluctuations of heavy-light modes in the momentum space are derived by taking a second derivative with respect to $\tilde{\Phi}$ in Eq.~\eqref{GammaEx}.
When we take the direction of the QCD Kondo exciton's propagations along the third axis, we find that four independent channels of $(\sigma_a,V_a^3)$, $(\pi_a,A^3_a)$, ($V^1_a.A^2_a$), and $(V^2_a,A^1_a)$ exist due to the parity invariance.
The energy and the three-dimensional momentum for the QCD Kondo excitons are expressed by $q_{0}$ and ${\bm q}$, respectively.
After some calculations, in each channel, we obtain the inverse propagators for $a=1$, which corresponds to {\it the dressed Kondo excitons},
\begin{widetext}
\begin{align}
{\cal D}_{\sigma_1 V_1^3}^{-1}= \left(
\begin{array}{cccc}
\bar{D}_{\sigma_1^\dagger\sigma_1} & \bar{D}_{\sigma_1^\dagger V_1^3} & \bar{D}_{\sigma_1^\dagger \sigma_1^\dagger}  & \bar{D}_{\sigma_1^\dagger V_1^{3\dagger}} \\
\bar{D}_{V_1^{3\dagger}\sigma_1} & \bar{D}_{V_1^{3\dagger} V_1^3} &\bar{D}_{V_1^{3\dagger} \sigma_1^\dagger} & \bar{D}_{V_1^{3\dagger} V_1^{3\dagger}} \\
\bar{D}_{\sigma_1\sigma_1}  & \bar{D}_{\sigma_1 V_1^3}  & \bar{D}_{\sigma_1 \sigma_1^\dagger} & \bar{D}_{\sigma_1 V_1^{3\dagger}} \\
\bar{D}_{V_1^{3}\sigma_1} & \bar{D}_{V_1^{3} V_1^3}  & \bar{D}_{V_1^{3} \sigma_1^\dagger}  & \bar{D}_{V_1^{3} V_1^{3\dagger}} \\
\end{array}
\right)  = \left(
\begin{array}{cccc}
{\cal I}_2& {\cal I}_9 & {\cal I}_1  & {\cal I}_7 \\
{\cal I}_9 &{\cal I}_2 & {\cal I}_8  & {\cal I}_5 \\
{\cal I}_1  & {\cal I}_8 & {\cal I}_3 & {\cal I}_{10} \\
{\cal I}_{7}  & {\cal I}_5 & {\cal I}_{10}  & {\cal I}_3 \\
\end{array}
\right) , \label{SigmaV}
\end{align}
\begin{align}
{\cal D}_{\pi_1 A_1^3}^{-1}= \left(
\begin{array}{cccc}
\bar{D}_{\pi_1^\dagger\pi_1} & \bar{D}_{\pi_1^\dagger A_1^3} & \bar{D}_{\pi_1^\dagger \pi_1^\dagger}  & \bar{D}_{\pi_1^\dagger A_1^{3\dagger}} \\
\bar{D}_{A_1^{3\dagger}\pi_1} & \bar{D}_{A_1^{3\dagger} A_1^3} &\bar{D}_{A_1^{3\dagger} \pi_1^\dagger} & \bar{D}_{A_1^{3\dagger} A_1^{3\dagger}} \\
\bar{D}_{\pi_1\pi_1}  & \bar{D}_{\pi_1 A_1^3}  & \bar{D}_{\pi_1 \pi_1^\dagger} & \bar{D}_{\pi_1 A_1^{3\dagger}} \\
\bar{D}_{A_1^{3}\pi_1} & \bar{D}_{A_1^{3} A_1^3}  & \bar{D}_{A_1^{3} \pi_1^\dagger}  & \bar{D}_{A_1^{3} A_1^{3\dagger}} \\
\end{array}
\right) =  \left(
\begin{array}{cccc}
{\cal I}_2  & -i{\cal I}_9 & {\cal I}_6 & -i{\cal I}_8 \\
i{\cal I}_9 &  {\cal I}_2  & -i{\cal I}_7  & {\cal I}_1 \\
{\cal I}_6 & i{\cal I}_7 & {\cal I}_3 & i{\cal I}_{10} \\
i{\cal I}_8 & {\cal I}_1  &-i{\cal I}_{10} & {\cal I}_3\\
\end{array}
\right)  , \label{PiA}
\end{align}
\begin{align}
{\cal D}_{V_1^1 A_1^2}^{-1}= \left(
\begin{array}{cccc}
\bar{D}_{V_1^{1\dagger} V_1^1} & \bar{D}_{V_1^{1\dagger} A_1^2} & \bar{D}_{V_1^{1\dagger} V_1^{1\dagger}  } & \bar{D}_{V_1^{1\dagger} A_1^{2\dagger}} \\
\bar{D}_{A_1^{2\dagger}V_1^1} & \bar{D}_{A_1^{2\dagger} A_1^2} &\bar{D}_{A_1^{2\dagger} V_1^{1\dagger}} & \bar{D}_{A_1^{2\dagger} A_1^{2\dagger}} \\
\bar{D}_{V_1^1V^1_1}  & \bar{D}_{V^1_1 A_1^2}  & \bar{D}_{V_1^1 V_1^{1\dagger}} & \bar{D}_{V_1^1 A_1^{2\dagger}} \\
\bar{D}_{A_1^{2}V_1^1} & \bar{D}_{A_1^{2} A_1^2}  & \bar{D}_{A_1^{2} V_1^{1\dagger}}  & \bar{D}_{A_1^{2} A_1^{2\dagger}} \\
\end{array}
\right)= \left(
\begin{array}{cccc}
{\cal I}_2 & i{\cal I}_9  & {\cal I}_4 & i{\cal I}_8 \\
-i{\cal I}_9  & {\cal I}_2& i{\cal I}_7  & {\cal I}_1 \\
{\cal I}_4  &-i{\cal I}_7 & {\cal I}_3  &-i{\cal I}_{10} \\
-i{\cal I}_8 &{\cal I}_1  & i{\cal I}_{10} & {\cal I}_3 \\
\end{array}
\right)  , \label{V1A2}
\end{align}
\begin{align}
{\cal D}_{V_1^2 A_1^1}^{-1}= \left(
\begin{array}{cccc}
\bar{D}_{V_1^{2\dagger} V_1^2} & \bar{D}_{V_1^{2\dagger} A_1^1} & \bar{D}_{V_1^{2\dagger} V_1^{2\dagger}  } & \bar{D}_{V_1^{2\dagger} A_1^{1\dagger}} \\
\bar{D}_{A_1^{1\dagger}V_1^2} & \bar{D}_{A_1^{1\dagger} A_1^1} &\bar{D}_{A_1^{1\dagger} V_1^{2\dagger}} & \bar{D}_{A_1^{1\dagger} A_1^{1\dagger}} \\
\bar{D}_{V_1^2V^2_1}  & \bar{D}_{V^2_1 A_1^1}  & \bar{D}_{V_1^2 V_1^{2\dagger}} & \bar{D}_{V_1^2 A_1^{1\dagger}} \\
\bar{D}_{A_1^{1}V_1^2} & \bar{D}_{A_1^{1} A_1^1}  & \bar{D}_{A_1^{1} V_1^{2\dagger}}  & \bar{D}_{A_1^{1} A_1^{1\dagger}} \\
\end{array}
\right)  =  \left(
\begin{array}{cccc}
{\cal I}_2 & -i{\cal I}_9& {\cal I}_4  & -i{\cal I}_8 \\
i{\cal I}_9 & {\cal I}_2  &-i{\cal I}_7& {\cal I}_1 \\
{\cal I}_4 & i{\cal I}_7 &{\cal I}_3  &i{\cal I}_{10} \\
i{\cal I}_8& {\cal I}_1  & -i{\cal I}_{10}  &{\cal I}_3 \\
\end{array}
\right)  , \label{V2A1}
\end{align}
with the functions defined by
\begin{eqnarray}
{\cal I}_1 &\equiv& \frac{2|\Delta|^2}{(2\pi)^2}\int_0^\Lambda dk\, k^2\int_{-1}^1dt\Bigg\{\frac{1}{(E_{k_-}^+-E_{k_-}^-)(q_0+E_{k_-}^+-E_{k_+}^+)(q_0+E_{k_-}^+-E_{k_+}^--i\eta)} \nonumber\\
&&+\frac{1}{(q_0-E_{k_+}^++E_{k_-}^+)(q_0-E_{k_+}^++E_{k_-}^-+i\eta)(E_{k_+}^+-E_{k_+}^-)}\Bigg\}\ ,
\end{eqnarray}
\begin{eqnarray}
{\cal I}_2 &\equiv& \frac{2}{(2\pi)^2}\int_0^\Lambda dk\, k^2 \int_{-1}^1dt\Bigg\{\frac{\Big((E_{k_-}^++\mu)E_{k_-}^+-|\Delta|^2\Big)(q_0+E_{k_-}^+-E_{k_+})}{(E_{k_-}^+-\tilde{E}_{k_-})(E_{k_-}^+-E_{k_-}^-)(q_0+E_{k_-}^+-E_{k_+}^+)(q_0+E_{k_-}^+-E_{k_+}^--i\eta)} \nonumber\\
&&-\frac{\Big((q_0-E_{k_+}^+-\mu)(q_0-E_{k_+}^+)-\Delta^2\Big)(E_{k_+}^+-E_{k_+})}{(q_0-E_{k_+}^++\tilde{E}_{k_-}+i\eta)(q_0-E_{k_+}^++E_{k_-}^+)(q_0-E_{k_+}^++E_{k_-}^-+i\eta)(E_{k_+}^+-E_{k_+}^-)}\Bigg\}-\frac{1}{G}\ ,
\end{eqnarray}
\begin{eqnarray}
{\cal I}_3 &\equiv& \frac{2}{(2\pi)^2}\int_0^\Lambda dk\, k^2 \int_{-1}^1dt\Bigg\{\frac{\Big((q_0+E_{k_-}^++\mu)(q_0+E_{k_-}^+)-\Delta^2\Big)(E_{k_-}^+-E_{k_-})}{(E_{k_-}^+-E_{k_-}^-)(q_0+E_{k_-}^+-\tilde{E}_{k_+}-i\eta)(q_0+E_{k_-}^+-E_{k_+}^+)(q_0+E_{k_-}^+-E_{k_+}^--i\eta)} \nonumber\\
&&-\frac{\Big((E_{k_+}^++\mu)E_{k_+}^+-\Delta^2\Big)(q_0-E_{k_+}^++E_{k_-})}{(q_0-E_{k_+}^++E_{k_-}^+)(q_0-E_{k_+}^++E_{k_-}^-+i\eta)(E_{k_+}^+-\tilde{E}_{k_+})(E_{k_+}^+-E_{k_+}^-)} \Bigg\}-\frac{1}{G}\ ,
\end{eqnarray}
\begin{eqnarray}
{\cal I}_4 &\equiv& \frac{2|\Delta|^2}{(2\pi)^2}\int_0^\Lambda  dk\, k^2\int_{-1}^1dt\Bigg\{\frac{\frac{1}{k_+k_-}\left(-k^2t^2+\frac{q^2}{4}\right)}{(E_{k_-}^+-E_{k_-}^-)(q_0+E_{k_-}^+-E_{k_+}^+)(q_0+E_{k_-}^+-E_{k_+}^--i\eta)} \nonumber\\
&&+\frac{\frac{1}{k_+k_-}\left(-k^2t^2+\frac{q^2}{4}\right)}{(q_0-E_{k_+}^++E_{k_-}^+)(q_0-E_{k_+}^++E_{k_-}^-+i\eta)(E_{k_+}^+-E_{k_+}^-)}\Bigg\}\ ,
\end{eqnarray}
\begin{eqnarray}
{\cal I}_5 &\equiv& \frac{2|\Delta|^2}{(2\pi)^2}\int_0^\Lambda dk\, k^2\int_{-1}^1dt\Bigg\{\frac{\frac{1}{k_+k_-}\left((2t^2-1)k^2-\frac{q^2}{4}\right)}{(E_{k_-}^+-E_{k_-}^-)(q_0+E_{k_-}^+-E_{k_+}^+)(q_0+E_{k_-}^+-E_{k_+}^--i\eta)} \nonumber\\
&&+\frac{\frac{1}{k_+k_-}\left((2t^2-1)k^2 - \frac{q^2}{4}\right)}{(q_0-E_{k_+}^++E_{k_-}^+)(q_0-E_{k_+}^++E_{k_-}^-+i\eta)(E_{k_+}^+-E_{k_+}^-)}\Bigg\}\ ,
\end{eqnarray}
\begin{eqnarray}
{\cal I}_6 &\equiv& - \frac{2|\Delta|^2}{(2\pi)^2}\int_0^\Lambda dk\, k^2\int_{-1}^1dt\Bigg\{\frac{\frac{1}{k_+k_-}\left(k^2-\frac{q^2}{4}\right)}{(E_{k_-}^+-E_{k_-}^-)(q_0+E_{k_-}^+-E_{k_+}^+)(q_0+E_{k_-}^+-E_{k_+}^--i\eta)} \nonumber\\
&&+\frac{\frac{1}{k_+k_-}\left(k^2-\frac{q^2}{4}\right)}{(q_0-E_{k_+}^++E_{k_-}^+)(q_0-E_{k_+}^++E_{k_-}^-+i\eta)(E_{k_+}^+-E_{k_+}^-)}\Bigg\}\ ,
\end{eqnarray}
\begin{eqnarray}
{\cal I}_7 &\equiv& \frac{2|\Delta|^2}{(2\pi)^2}\int_0^\Lambda dk\, k^2\int_{-1}^1dt\Bigg\{\frac{\frac{1}{k_+}\left(kt+\frac{q}{2}\right)}{(E_{k_-}^+-E_{k_-}^-)(q_0+E_{k_-}^+-E_{k_+}^+)(q_0+E_{k_-}^+-E_{k_+}^--i\eta)} \nonumber\\
&&+\frac{\frac{1}{k_+}\left(kt+\frac{q}{2}\right)}{(q_0-E_{k_+}^++E_{k_-}^+)(q_0-E_{k_+}^++E_{k_-}^-+i\eta)(E_{k_+}^+-E_{k_+}^-)}\Bigg\}\ ,
\end{eqnarray}
\begin{eqnarray}
{\cal I}_8 &\equiv& \frac{2|\Delta|^2}{(2\pi)^2}\int_0^\Lambda dk\, k^2\int_{-1}^1dt\Bigg\{\frac{\frac{1}{k_-}\left(kt-\frac{q}{2}\right)}{(E_{k_-}^+-E_{k_-}^-)(q_0+E_{k_-}^+-E_{k_+}^+)(q_0+E_{k_-}^+-E_{k_+}^--i\eta)} \nonumber\\
&&+\frac{\frac{1}{k_-}\left(kt-\frac{q}{2}\right)}{(q_0-E_{k_+}^++E_{k_-}^+)(q_0-E_{k_+}^++E_{k_-}^-+i\eta)(E_{k_+}^+-E_{k_+}^-)}\Bigg\}\ ,
\end{eqnarray}
\begin{eqnarray}
{\cal I}_9 &\equiv& \frac{2}{(2\pi)^2}\int_0^\Lambda dk\, k^2\int_{-1}^1dt\Bigg\{\frac{\Big(k_-E_{k_-}^++|\Delta|^2\Big)(q_0+E_{k_-}^+-E_{k_+})\frac{1}{k_-}\left(kt-\frac{q}{2}\right)}{(E_{k_-}^+-\tilde{E}_{k_-})(E_{k_-}^+-E_{k_-}^-)(q_0+E_{k_-}^+-E_{k_+}^+)(q_0+E_{k_-}^+-E_{k_+}^--i\eta)} \nonumber\\
&&+\frac{\Big(k_-(q_0-E_{k_+}^+)-|\Delta|^2\Big)(E_{k_+}^+-E_{k_+})\frac{1}{k_-}\left(kt-\frac{q}{2}\right)}{(q_0-E_{k_+}^++\tilde{E}_{k_-}+i\eta)(q_0-E_{k_+}^++E_{k_-}^+)(q_0-E_{k_+}^++E_{k_-}^-+i\eta)(E_{k_+}^+-E_{k_+}^-)}\Bigg\} \ ,
\end{eqnarray}
\begin{eqnarray}
{\cal I}_{10} &\equiv& \frac{2}{(2\pi)^2}\int_0^\Lambda dk\, k^2\int_{-1}^1dt\Bigg\{\frac{\Big(k_+(q_0+E_{k_-}^+)+|\Delta|^2\Big)(E_{k_-}^+-E_{k_-})\frac{1}{k_+}\left(kt+\frac{q}{2}\right)}{(E_{k_-}^+-E_{k_-}^-)(q_0+E_{k_-}^+-\tilde{E}_{k_+}-i\eta)(q_0+E_{k_-}^+-E_{k_+}^+)(q_0+E_{k_-}^+-E_{k_+}^--i\eta)} \nonumber\\
&&-\frac{\Big(k_+E_{k_+}^++|\Delta|^2\Big)(q_0-E_{k_+}^++E_{k_-})\frac{1}{k_+}\left(kt+\frac{q}{2}\right)}{(q_0-E_{k_+}^++E_{k_-}^+)(q_0-E_{k_+}^++E_{k_-}^-+i\eta)(E_{k_+}^+-\tilde{E}_{k_+})(E_{k_+}^+-E_{k_+}^-)} \Bigg\} \ ,
\end{eqnarray}
where ${\bm k}_\pm = {\bm k}\pm{\bm q}/{2}$. In the calculations of the matrix elements, for example, $\bar{D}_{\sigma^\dagger_1\sigma_1}$ is defined by
\begin{align}
\frac{\delta^2\Gamma_{\rm eff}[\{\Phi\};\Delta]}{\delta \tilde{\sigma}_1^\dagger(p) \, \delta\tilde{\sigma}_1(q)}\Bigg|_{\tilde{\Phi}=0} = \bar{D}_{\sigma_1^\dagger\sigma_1}(q)(2\pi)^4\delta^4(p+q) .  \label{TwoPoint}
\end{align}
Similar equations hold also for the other channels.

For $a=2,3,\cdots, N_f$ for {\it the half-dressed Kondo excitons}, the quark inverse propagators, which are alternative to Eqs.~\eqref{SigmaV}-\eqref{V2A1}, are
\begin{align}
{\cal D}_{\sigma_a V_a^3}^{-1}= \left(
\begin{array}{cccc}
\bar{D}_{\sigma_a^\dagger\sigma_a} & \bar{D}_{\sigma_a^\dagger V_a^3} & \bar{D}_{\sigma_a^\dagger \sigma_a^\dagger}  & \bar{D}_{\sigma_a^\dagger V_a^{3\dagger}} \\
\bar{D}_{V_a^{3\dagger}\sigma_a} & \bar{D}_{V_a^{3\dagger} V_
a^3} &\bar{D}_{V_a^{3\dagger} \sigma_a^\dagger} & \bar{D}_{V_a^{3\dagger} V_a^{3\dagger}} \\
\bar{D}_{\sigma_a\sigma_a}  & \bar{D}_{\sigma_a V_a^3}  & \bar{D}_{\sigma_a \sigma_a^\dagger} & \bar{D}_{\sigma_a V_a^{3\dagger}} \\
\bar{D}_{V_a^{3}\sigma_a} & \bar{D}_{V_a^{3} V_a^3}  & \bar{D}_{V_a^{3} \sigma_a^\dagger}  & \bar{D}_{V_a^{3} V_a^{3\dagger}} \\
\end{array}
\right)   = \left(
\begin{array}{cccc}
{\cal J}_2& {\cal J}_9 & 0 & 0 \\
{\cal J}_9&{\cal J}_2  & 0  &0 \\
0  &0 & {\cal J}_3 & {\cal J}_{10} \\
0  & 0 & {\cal J}_{10}  & {\cal J}_3 \\
\end{array}
\right) , \label{SigmaVN}
\end{align}
\begin{align}
{\cal D}_{\pi_a A_a^3}^{-1}= \left(
\begin{array}{cccc}
\bar{D}_{\pi_a^\dagger\pi_a} & \bar{D}_{\pi_a^\dagger A_a^3} & \bar{D}_{\pi_a^\dagger \pi_a^\dagger}  & \bar{D}_{\pi_a^\dagger A_a^{3\dagger}} \\
\bar{D}_{A_a^{3\dagger}\pi_a} & \bar{D}_{A_a^{3\dagger} A_a^3} &\bar{D}_{A_a^{3\dagger} \pi_a^\dagger} & \bar{D}_{A_a^{3\dagger} A_a^{3\dagger}} \\
\bar{D}_{\pi_a\pi_a}  & \bar{D}_{\pi_a A_a^3}  & \bar{D}_{\pi_a \pi_a^\dagger} & \bar{D}_{\pi_a A_a^{3\dagger}} \\
\bar{D}_{A_a^{3}\pi_a} & \bar{D}_{A_a^{3} A_a^3}  & \bar{D}_{A_a^{3} \pi_a^\dagger}  & \bar{D}_{A_a^{3} A_a^{3\dagger}} \\
\end{array}
\right) =  \left(
\begin{array}{cccc}
{\cal J}_2  & -i{\cal J}_9 &0 & 0 \\
i{\cal J}_9 &  {\cal J}_2  & 0  & 0 \\
0 & 0 & {\cal J}_3 & i{\cal J}_{10} \\
0 & 0 &  -i{\cal J}_{10}  & {\cal J}_3\\
\end{array}
\right)   , \label{PiAN}
\end{align}
\begin{align}
{\cal D}_{V_a^1 A_a^2}^{-1}= \left(
\begin{array}{cccc}
\bar{D}_{V_a^{1\dagger} V_a^1} & \bar{D}_{V_a^{1\dagger} A_a^2} & \bar{D}_{V_a^{1\dagger} V_a^{1\dagger}  } & \bar{D}_{V_a^{1\dagger} A_a^{2\dagger}} \\
\bar{D}_{A_a^{2\dagger}V_a^1} & \bar{D}_{A_a^{2\dagger} A_a^2} &\bar{D}_{A_a^{2\dagger} V_a^{1\dagger}} & \bar{D}_{A_a^{2\dagger} A_a^{2\dagger}} \\
\bar{D}_{V_a^1V^1_a}  & \bar{D}_{V^1_a A_a^2}  & \bar{D}_{V_a^1 V_a^{1\dagger}} & \bar{D}_{V_a^1 A_a^{2\dagger}} \\
\bar{D}_{A_a^{2}V_a^1} & \bar{D}_{A_a^{2} A_a^2}  & \bar{D}_{A_a^{2} V_a^{1\dagger}}  & \bar{D}_{A_a^{2} A_a^{2\dagger}} \\
\end{array}
\right)= \left(
\begin{array}{cccc}
{\cal J}_2   & i{\cal J}_9 &0 & 0\\
-i{\cal J}_9 & {\cal J}_2  &0 &0\\
0  &0 & {\cal J}_3 &-i{\cal J}_{10} \\
0  & 0 & i{\cal J}_{10} & {\cal J}_3 \\
\end{array}
\right)  , \label{V1A2N}
\end{align}
\begin{align}
{\cal D}_{V_a^2 A_a^1}^{-1}= \left(
\begin{array}{cccc}
\bar{D}_{V_a^{2\dagger} V_a^2} & \bar{D}_{V_a^{2\dagger} A_a^1} & \bar{D}_{V_a^{2\dagger} V_a^{2\dagger}  } & \bar{D}_{V_a^{2\dagger} A_a^{1\dagger}} \\
\bar{D}_{A_a^{1\dagger}V_a^2} & \bar{D}_{A_a^{1\dagger} A_a^1} &\bar{D}_{A_a^{1\dagger} V_a^{2\dagger}} & \bar{D}_{A_a^{1\dagger} A_a^{1\dagger}} \\
\bar{D}_{V_a^2V^2_a}  & \bar{D}_{V^2_a A_a^1}  & \bar{D}_{V_a^2 V_a^{2\dagger}} & \bar{D}_{V_a^2 A_a^{1\dagger}} \\
\bar{D}_{A_a^{1}V_a^2} & \bar{D}_{A_a^{1} A_a^1}  & \bar{D}_{A_a^{1} V_a^{2\dagger}}  & \bar{D}_{A_a^{1} A_a^{1\dagger}} \\
\end{array}
\right)  =  \left(
\begin{array}{cccc}
{\cal J}_2 & -i{\cal J}_9 & 0  & 0 \\
i{\cal J}_9  & {\cal J}_2 &0 & 0 \\
0&0  &{\cal J}_3  &i{\cal J}_{10} \\
0  &0& -i{\cal J}_{10} &{\cal J}_3 \\
\end{array}
\right)  , \label{V2A1N}
\end{align}
with the functions defined by
\begin{eqnarray}
{\cal J}_2 &\equiv& \frac{2}{(2\pi)^2} \int_0^\Lambda dk\, k^2\int_{-1}^1dt\Bigg\{ \frac{-(q_0-E_{k_+}^+-\mu)(E_{k_+}^+-E_{k_+})}{(q_0-E_{k_+}^++\tilde{E}_{k_-}+i\eta)(q_0-E_{k_+}^++E_{k_-}+i\theta(\mu-k_-)\eta)(E_{k_+}^+-E_{k_+}^-)} \nonumber\\
&& + \frac{\theta(k_--\mu)(E_{k_-}+\mu)(q_0+E_{k_-}-E_{k_+})}{(E_{k_-}-\tilde{E}_{k_-})(q_0+E_{k_-}-E_{k_+}^+)(q_0+E_{k_-}-E_{k_+}^--i\eta)} \Bigg\} - \frac{1}{G}\ ,
\end{eqnarray}
\begin{eqnarray}
{\cal J}_3 &\equiv& \frac{2}{(2\pi)^2} \int_0^\Lambda dk\, k^2\int_{-1}^1dt\Bigg\{ \frac{(q_0+E_{k_-}^++\mu)(E_{k_-}^+-E_{k_-})}{(q_0+E_{k_-}^+-\tilde{E}_{k_+}-i\eta)(q_0+E_{k_-}^+-E_{k_+}-i\theta(\mu-k_+)\eta)(E_{k_-}^+-E_{k_-}^-)} \nonumber\\
&& - \frac{\theta(k_+-\mu)(E_{k_+}+\mu)(q_0-E_{k_+}+E_{k_-})}{(E_{k_+}-\tilde{E}_{k_+})(q_0-E_{k_+}+E_{k_-}^+)(q_0-E_{k_+}+E_{k_-}^-+i\eta)} \Bigg\}- \frac{1}{G}\ ,
\end{eqnarray}
\begin{eqnarray}
{\cal J}_9 &\equiv& \frac{2}{(2\pi)^2} \int_0^\Lambda dk\, k^2\int_{-1}^1dt\Bigg\{ \frac{(E_{k_+}^+-E_{k_+})\left(kt-\frac{q}{2}\right)}{(q_0-E_{k_+}^++\tilde{E}_{k_-}+i\eta)(q_0-E_{k_+}^++E_{k_-}+i\theta(\mu-k_-)\eta)(E_{k_+}^+-E_{k_+}^-)} \nonumber\\
&& + \frac{\theta(k_--\mu)(q_0+E_{k_-}-E_{k_+})\left(kt-\frac{q}{2}\right)}{(E_{k_-}-\tilde{E}_{k_-})(q_0+E_{k_-}-E_{k_+}^+)(q_0+E_{k_-}-E_{k_+}^--i\eta)}\Bigg\} \ ,
\end{eqnarray}
\begin{eqnarray}
{\cal J}_{10} &\equiv& \frac{2}{(2\pi)^2} \int_0^\Lambda dk\, k^2\int_{-1}^1dt
\Bigg\{ \frac{(E_{k_-}^+-E_{k_-})\left(kt+\frac{q}{2}\right)}{(q_0+E_{k_-}^+-\tilde{E}_{k_+}-i\eta)(q_0+E_{k_-}^+-E_{k_+}-i\theta(\mu-k_+)\eta)(E_{k_-}^+-E_{k_-}^-)} \nonumber\\
&& - \frac{\theta(k_+-\mu)(q_0-E_{k_+}+E_{k_-})\left(kt+\frac{q}{2}\right)}{(E_{k_+}-\tilde{E}_{k_+})(q_0-E_{k_+}+E_{k_-}^+)(q_0-E_{k_+}+E_{k_-}^-+i\eta)}\Bigg\} \ .
\end{eqnarray}
The channels of $(\sigma_a,V_a^3)$, $(\pi_a,A^3_a)$, ($V^1_a.A^2_a$), and $(V^2_a,A^1_a)$ are independent of each other:
The energy-momentum dispersion relations for the dressed Kondo excitons are determined by solving ${\rm det}\bigl({\cal D}_{\sigma_1V_1^3}^{-1}\bigr)=0$, ${\rm det}\bigl({\cal D}_{\pi_1A_1^3}^{-1}\bigr)=0$, ${\rm det}\bigl({\cal D}_{V_1^1A_1^2}^{-1}\bigr)=0$, and ${\rm det}\bigl({\cal D}_{V_1^2A_1^1}^{-1}\bigr)=0$, while those for the half-dressed Kondo excitons are determined by ${\rm det}\bigl({\cal D}_{\sigma_aV_a^3}^{-1}\bigr)=0$, ${\rm det}\bigl({\cal D}_{\pi_aA_a^3}^{-1}\bigr)=0$, ${\rm det}\bigl({\cal D}_{V_a^1A_a^2}^{-1}\bigr)=0$, and ${\rm det}\bigl({\cal D}_{V_a^2A_a^1}^{-1}\bigr)=0$ ($a=2,3,\cdots, N_f$).
\end{widetext}

\bibliography{reference}

\end{document}